\documentclass[prl,amsmath,amssymb,twocolumn,showpacs]{revtex4}

\usepackage{graphicx}
\usepackage{subfigure}
\usepackage{adjustbox}
\usepackage{bm}
\usepackage{color}
\usepackage{braket}
\usepackage{standalone}
\usepackage{multirow}
\usepackage{tikz}
\usepackage{mathrsfs}
\usepackage{dsfont}
\usepackage[colorlinks,bookmarks=true,citecolor=blue,linkcolor=red,urlcolor=blue]{hyperref}

\begin{document}
\title{Biorthogonal Bulk-Boundary Correspondence in Non-Hermitian Systems
}

\author{Flore K. Kunst$^{1,*}$, Elisabet Edvardsson$^{1}$, Jan Carl Budich$^{2}$, and Emil J. Bergholtz$^{1,*}$}

\affiliation{$^1$ Department of Physics, Stockholm University, AlbaNova University Center, 106 91 Stockholm, Sweden
\\$^2$ Institute of Theoretical Physics, Technische Universit\"{a}t Dresden, 01062 Dresden, Germany}
\email[Corresponding author. ]{emil.bergholtz@fysik.su.se}
\email[Corresponding author. ]{f.k.kunst@gmail.com}
\date{\today}

\begin{abstract}

Non-Hermitian systems exhibit striking exceptions from the paradigmatic bulk-boundary correspondence, including the failure of bulk Bloch band invariants in predicting boundary states and the (dis)appearance of boundary states at parameter values far from those corresponding to gap closings in periodic systems without boundaries. Here, we provide a comprehensive framework to unravel this disparity based on the notion of biorthogonal quantum mechanics: While the properties of the left and right eigenstates corresponding to boundary modes are individually decoupled from the bulk physics in non-Hermitian systems, their combined biorthogonal density penetrates the bulk precisely when phase transitions occur. This leads to generalized bulk-boundary correspondence and a quantized biorthogonal polarization that is formulated directly in systems with open boundaries. We illustrate our general insights by deriving the phase diagram for several microscopic open boundary models, including exactly solvable non-Hermitian extensions of the Su-Schrieffer-Heeger model and Chern insulators.
\end{abstract}

\maketitle
The quest for a complete classification and comprehensive physical understanding of topological phases of matter such as topological insulators \cite{hasankane, qizhang, bansillindas} has been at the forefront of research in physics for many years. 
For closed systems, as described by Hermitian Hamiltonians, the so-called bulk-boundary correspondence represents a ubiquitous guiding principle to the phenomenology of topological insulators: bulk topological invariants characterizing a given phase are uniquely reflected in gapless (metallic) surface states.
This general pattern is found throughout the recently established hierarchy of topological phases, where $n$th order phases in $d$ spatial dimensions feature $d-n$ dimensional generalized boundary states \cite{benalcazarbernevighughes, parameswaranwan,langhehnpentrifuoppenbrouwer, songfangfang, schindlercookvergnio}. 

In contrast to this systematic picture, quite basic questions have so far remained unanswered for open systems governed by {\emph{non-Hermitian}} Hamiltonians \cite{reviewTorres} with applications ranging from various mechanical and optical meta-materials subject to gain and loss terms \cite{lujoannopoulossoljacic}, to quasiparticles with finite lifetime in heavy-fermion systems \cite{koziifu, yoshidapeterskawakmi}. The topological properties of such systems can crucially differ from their Hermitian counterparts \cite{esakisatohasebekohmoto, yuce, yuce1, harterleejoglekar, zhuluchen, xiong,wangzhangsong, lee, lieu, yinjiangliluchen, yaowang, shenzhenfu, yaosongwang, rudnerlevitov, alvarezvargastorres, leykambliokhhuangchongnori, wiemannkremerplotniklumernoltemakrissegevrechtsmanszameit,schomerus,reviewTorres}, as exemplified by the prediction and experimental observation of unconventional topological boundary modes in certain parity time-reversal ($\mathcal{P}\mathcal{T}$) symmetric systems \cite{benderboettcher, bender, yuce, yuce1, harterleejoglekar, wiemannkremerplotniklumernoltemakrissegevrechtsmanszameit}. At a more fundamental level, the generalization to non-Hermitian systems of the bulk-boundary correspondence is an open question that has recently become a subject of active and controversial discussion \cite{lee, xiong, yaowang, yaosongwang,leykambliokhhuangchongnori,reviewTorres}. The paramount issue is the so-called non-Hermitian skin effect, meaning that a macroscopic number of left and right eigenmodes can pile up exponentially close to the boundary \cite{xiong,alvarezvargastorres,yaowang}. In this scenario, where even the bulk spectra qualitatively depend on the boundary conditions, the conventional approach of predicting boundary modes from topological invariants defined for periodic systems does not provide a conclusive physical picture.

In this Letter, we overcome these issues by introducing a widely applicable approach to understand the enigmatic bulk-boundary correspondence in non-Hermitian systems using the framework of biorthogonal quantum mechanics \cite{brody}. 
There, a biorthogonal set of right and left eigenstates of a non-Hermitian Hamiltonian replaces the notion of a conventional eigensystem familiar from the Hermitian realm.
This generalization accounts for the fact that non-Hermitian Hamiltonians may feature a complex spectrum and non-orthogonal eigenstates.
Below, we construct a {\emph{biorthogonal polarization}} $P$ that is able to accurately predict the occurrence of boundary and domain-wall modes in a broad class of non-Hermitian systems, even in the presence of the aforementioned non-Hermitian skin effect. 
Quite remarkably, dealing with open boundary systems from the outset, our theory explains a number of intriguing numerical observations in a natural and unified way, such as the occurrence of boundary zeromodes in parameter regimes that are in striking disagreement with predictions from the conventional bulk-boundary correspondence. 
To elucidate and practically apply our theoretical analysis, we generalize an existing approach for analytically finding exact boundary states \cite{kunsttrescherbergholtz, kunstvmiertbergholtz} to a large set of non-Hermitian models including Su-Schrieffer-Heeger (SSH) chains \cite{ssh, zhuluchen, lee, lieu, yinjiangliluchen, yaowang} featuring zero modes at the ends, and the Rice-Mele model \cite{ricemele}, which realizes a non-Hermitian generalization of a Chern insulating phase with chiral edge states \cite{klitzingdordapepper, haldane, tknn}.

{\it Exact boundary modes in non-Hermitian models.---} We study generic lattice models that consist of two single-orbital sublattices, $A$ and $B$. For concreteness, we start with systems in one spatial dimension (1D), noting that examples of 2D models will also be discussed further below. The non-Hermitian Bloch Hamiltonian of a lattice with translational invariance then reads $H_k = \Phi_{k}^\dagger \mathcal{H}_k \Phi_{k}$, where $\Phi^\dagger_{k} = (c^\dagger_{A,k}, \, c^\dagger_{B, k})$ with $c^\dagger_{A(B),k}$ the creation operator of an electron on site $A$ ($B$) and
\begin{equation}
\mathcal{H}_k = {\bf d}(k) \cdot \boldsymbol\sigma, \qquad {\bf d}(k) \in \mathds{C}^3, \label{eqgenblochham}
\end{equation}
with $\boldsymbol\sigma$ the vector of Pauli matrices, and $k$ the quasimomentum.
Writing ${\bf d}_1 (k) \equiv {\rm Re}[{\bf d}(k)]$ and ${\bf d}_2 (k) \equiv {\rm Im}[{\bf d}(k)]$, the eigenvalues are $E_\pm(k) = \pm \sqrt{{\bf d}_1^2(k) - {\bf d}_2^2(k) + 2i \,{\bf d}_1(k) \cdot {\bf d}_2(k)}$, and the exceptional points (EPs) are retrieved by finding the degeneracy in the spectrum, i.e., $E_+(k)= E_- (k)= 0$. For a system with open boundaries whose real-space Hamiltonian is $H = \Phi^\dagger \mathcal{H} \Phi$, where $\Phi^\dagger = (c^\dagger_{A,1}, \, c^\dagger_{B,1}, \, c^\dagger_{A,2}, \, \ldots)$ with $c^\dagger_{A(B), n}$ the creation operator of an electron on sublattice $A$ ($B$) in unit cell $n$. The eigenvalue equations read
\begin{equation}
H \ket{\Psi_{R,i}} = E_{i} \ket{\Psi_{R,i}}, \quad H^\dagger \ket{\Psi_{L,i}} = E_i^* \ket{\Psi_{L,i}}, \label{eqsimplifiedschrodingereq}
\end{equation}
where $\ket{\Psi_{L(R),i}}$ are the left (right) eigenvectors with $i$ the band index.

We now proceed by considering models with short-range hopping defining $d_x (k) \pm i d_y (k) \equiv f_\pm + g_\pm {\rm e}^{\pm i k \cdot a}$, where $a$ is the lattice vector between neighboring unit cells. When the open system terminates with an $A$ site at both ends, i.e. the total number of sites is odd and the last unit cell is broken, we find that there is always a mode with energy $d_z$, whose wave function has an exactly disappearing amplitude on all $B$ sites due to destructive interference, such that
\begin{align}
\ket{\psi_R} &= \mathcal{N}_R \sum_{n=1}^N r_R^{n} c^\dagger_{A, n} \ket{0}, \qquad r_R  = - \frac{f_+}{g_+},\label{eqexactwavefctsolright} \\
\ket{\psi_L} &= \mathcal{N}_L \sum_{n=1}^N r_L^{n} c^\dagger_{A, n} \ket{0}, \qquad  r_L = - \frac{f_-^*}{g_-^*},\label{eqexactwavefctsolleft}
\end{align}
are exact eigenstates with $\mathcal{N}_{L(R)}$ the normalization factor, $\ket{\psi_{L(R)}} \in \{ \ket{\Psi_{L(R),i}}\}$ and $r_{L(R)}$ are found from solving Eq.~(\ref{eqsimplifiedschrodingereq}) for $\ket{\psi_{L(R)}}$ \cite{kunsttrescherbergholtz, kunstvmiertbergholtz}. Below we use these solutions to elucidate our general conclusions.

\begin{figure}[t]
  \centering
\includegraphics[width=0.99\columnwidth]{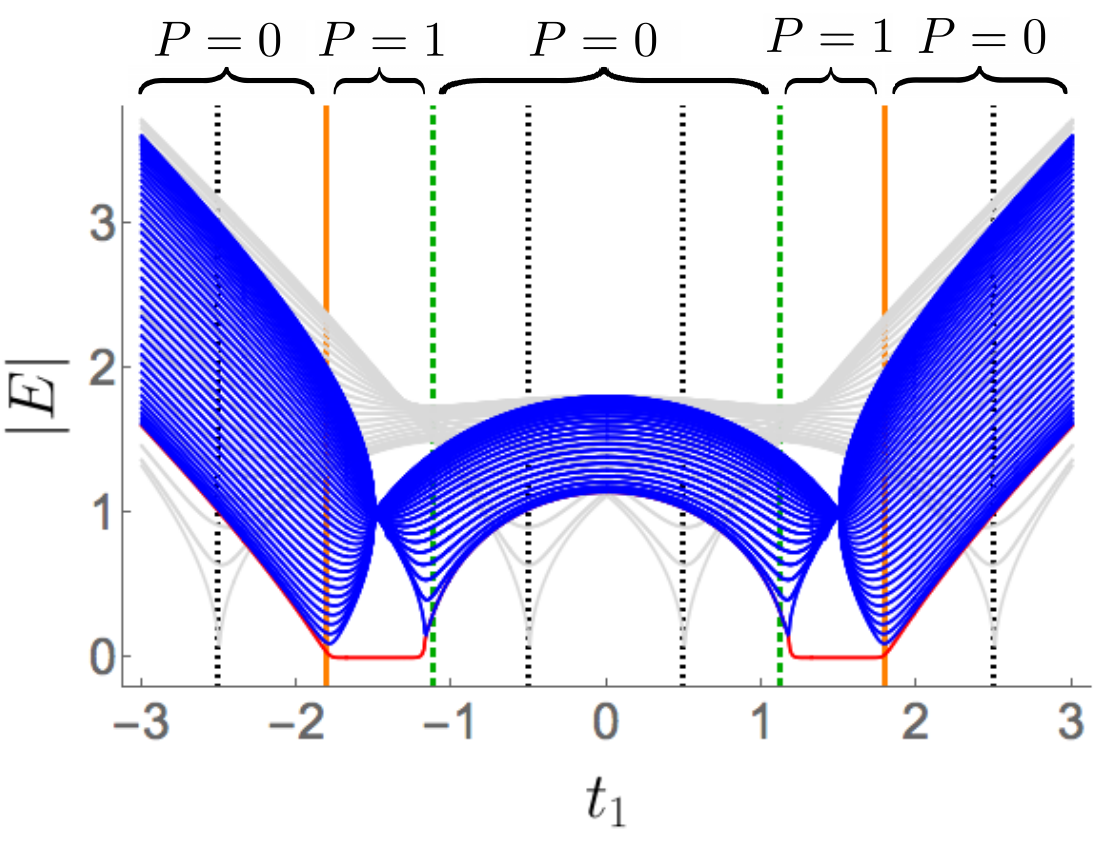}
  \caption{Energy spectra (absolute values) of the non-Hermitian SSH model for $t_1=t_2 = 1$, $\gamma = 3$ and $N = 46$. The gray lines indicate the periodic Bloch bands while the qualitatively different spectra in blue (bulk) and red (edge) correspond to the open system. The orange (dark green dashed) vertical lines indicate where $r_L^* r_R= 1$ ($r_L^* r_R= -1$) and the gray dotted-dashed lines correspond to the EPs of the periodic Bloch Hamiltonian. The value of $P$ is indicated by brackets and shows that we indeed find two zero modes when $P=1$, no zero modes when $P=0$, and that the bulk gap closes precisely where $P$ jumps (the small deviation is a finite size effect).}
  \label{fig_energy_semissh}
\end{figure}

{\it Biorthogonal bulk-boundary correspondence.---} In Hermitian systems, the occurrence of protected boundary modes can be determined from the bulk-phase diagram of a periodic system without boundaries, where transition points are marked by bulk-band touchings. In contrast, in non-Hermitian systems the bulk-band touchings might discontinuously shift in parameter space when going from periodic to open boundary conditions (e.g., see the qualitative difference between the gray and blue spectra in Fig.~\ref{fig_energy_semissh}). This phenomenon shows that the conventional bulk-boundary correspondence cannot be generally valid for non-Hermitian systems. Here, we remedy this issue by constructing an analogue of the bulk-boundary correspondence, where both the bulk and the edge quantities are defined in systems with open boundaries. We start by noting that, in sharp contrast to Hermitian matrices, the right and left eigenvectors of non-Hermitian Hamiltonians are in general different and not necessarily mutually orthogonal. To address this situation one can choose the normalization $\braket{\Psi_{L,i}|\Psi_{R,j}} = \delta_{i,j}$ to obtain a {\emph{biorthogonal}} set \cite{brody}, which is complete away from EPs, such that $\mathcal{N}_{L}^* \mathcal{N}_{R} = (r_L^* r_R)^{-1} (r_L^* r_R - 1)/[(r_L^* r_R)^N - 1]$ for the states described in Eqs.~(\ref{eqexactwavefctsolright}) and (\ref{eqexactwavefctsolleft}). To study the localization of modes in the biorthogonal context, we consider the biorthogonal expectation value of the projection operator $\Pi_n = \ket{e_{A, n}}\bra{e_{A, n}}+\ket{e_{B, n}}\bra{e_{B, n}}$ with $\ket{e_{A(B),n}} = c^\dagger_{A(B), n} \ket{0}$ onto unit cell $n$. 
For the exact solutions we find $\left<\Pi_n\right> \equiv  \left<\psi_L \left|\Pi_n\right|\psi_R\right> = \mathcal{N}_{L}^* \mathcal{N}_{R} (r_L^* r_R)^{n}$, such that $|r_L^* r_R|<1$ ($|r_L^* r_R|> 1$) means the mode is exponentially localized to the unit cell $n = 1$ ($n = N$), whereas Eqs.~(\ref{eqexactwavefctsolright}) and (\ref{eqexactwavefctsolleft}) describe an exact bulk state when
\begin{equation}
|r_L^* r_R|=1. \label{eqimportantsolv}
\end{equation}
This criticality condition correctly marks the boundary between parameter regions where the number and/or localization properties of boundary modes change. Note that in the Hermitian limit $r_R = r_L$, and Eq.~(\ref{eqimportantsolv}) still accurately predicts the (dis)appearance of boundary modes. More generally, for left and right boundary states with penetration length $\xi_L$ and $\xi_R$, respectively, i.e., $\left<n|\psi_{L(R)}\right>\sim e^{-n/\xi_{L(R)}}$ with $\ket{n} = \ket{e_{A,n}} + \ket{e_{B,n}}$, the corresponding condition reads 
\begin{equation}
\xi_R+\xi_L=0 , \label{eqimportant}
\end{equation}
which highlights the fact that, at criticality, a biorthogonal bulk state forms from right and left states localized at opposite ends.
 
Using the eigenstates $\ket{\psi_R}, \ket{\psi_L}$ we now construct a crucial quantity coined {\it biorthogonal polarization} $P$, which is defined as
\begin{equation}
P \equiv 1-\underset{N \rightarrow \infty}{{\rm lim}} \left<\psi_L \left|\frac{\sum_n n \, \Pi_n}{N}\right|\psi_R\right> 
 \label{eqbiorthpol}
\end{equation}
and exhibits a quantized jump precisely at the points determined by the condition (\ref{eqimportant}). While the biorthogonal density $\left<\Pi_n\right>$ is in general complex valued, $P$ is quantized for any boundary state independent of details by virtue of the biorthogonal normalization condition. Physically, $P$ is a quantitative measure for the biorthogonal localization of the mode which is towards $n=1$ ($n=N$) when $P=1$ ($P=0$). Accordingly, $P=1$ when $\frac 1{\xi_L} +\frac 1 {\xi_R}>0$ (or $|r_L^* r_R|<1$) and $P=0$ when $\frac 1{\xi_L} +\frac 1 {\xi_R}<0$ (or $|r_L^* r_R|>1$). This leads to a biorthogonal bulk-boundary correspondence in the following sense: when $P$ jumps between $0$ and $1$, the gap between the boundary state and the bulk spectrum must close in the {\emph{open system}}. It is crucial to note that a polarization calculated from either the right or left eigenstates alone would give different predictions that are not compatible with numerical observations on various systems. Here, instead we show explicitly in several examples that Eq.~(\ref{eqimportant}) accurately predicts band touchings associated with changes in the number and spatial localization of boundary modes at a given end of the system.

It is instructive to consider systems that are terminated in different ways. For this purpose, the solvable models are particularly illustrative: if the chain terminates with $B$ sites on both ends, the wave functions in Eqs.~(\ref{eqexactwavefctsolright}) and (\ref{eqexactwavefctsolleft}) are still exact with $A \rightarrow B$ and  $r_{L(R)} \rightarrow 1/r_{L(R)}$, such that the $P=0,1$ regions are interchanged while leaving the transition points unaltered. For a chain with an even number of sites, Eqs.~(\ref{eqexactwavefctsolright}) and (\ref{eqexactwavefctsolleft}) are no longer exact solutions, while their biorthogonal polarization $P$ remains the relevant predictive quantity: two modes localized at the ends of the chain appear when $P=1$ ($|r_L^* r_R| <1$), since both ends locally map onto the end $m=1$ of the odd chain, while for $P = 0$ ($|r_L^* r_R| \geq 1$) such a mapping entails no end modes. 

To illustrate these general results, we now discuss several concrete and instructive examples.

{\it Non-Hermitian SSH models.---} As a first example, we study a non-Hermitian version of the SSH model \cite{ssh}, where the Bloch Hamiltonian in Eq.~(\ref{eqgenblochham}) takes the concrete form \cite{yinjiangliluchen, yaowang}
\begin{equation}
d_x(k) = t_1 + t_2 \, {\rm cos}(k), \quad d_y(k) = i \frac{\gamma}{2} + t_2 \, {\rm sin}(k), \quad d_z = 0, \label{eqsshham}
\end{equation}
 which gives $f_\pm = t_1 \mp \gamma/2$ and $g_\pm = t_2$. This model has a chiral symmetry, $\sigma_z H_k \sigma_z = - H_k$, which means the eigenvalues appear in pairs $(E_n, -E_n)$ which is crucial for the appearance of protected zero-energy boundary modes both in Hermitian and in the present non-Hermitian systems. We find four EPs: $t_1 = \pm \gamma/2 - t_2$ at $k = 0$ and $t_1 = \pm \gamma/2 + t_2$ at $k = \pi$. For open boundary conditions, we find $r_R = - (t_1 - \gamma/2)/t_2$ and $r_L = - (t_1 + \gamma/2)/t_2$ leading to $r^*_L r_R = (t_1^2 - \gamma^2/4)/t_2^2$, and the bulk zero mode appears when Eq.~(\ref{eqimportant}) is satisfied, i.e.,
\begin{equation}
t_1 = \pm \sqrt{\frac{\gamma^2}{4} + t_2^2 }, \,\,\, \pm \sqrt{\frac{\gamma^2}{4} -t_2^2  }. \label{eqresultopenssh}
\end{equation}
By comparing the energy spectra with open (blue) and periodic (gray) boundaries for a chain with an integer number of unit cells in Fig.~\ref{fig_energy_semissh}, we observe a striking confirmation of our prediction: while the biorthogonal bulk-boundary correspondence holds true, the periodic spectra exhibit a gap closing at very different parameter values \cite{otherapproach}, thus highlighting the breakdown of the conventional bulk-boundary correspondence.
We note that, by coincidence, the EPs of the periodic system coincide with the points where $|r_R| = 1$ and $|r_L|=1$, although there is no such generic correspondence. We also note that the zero energy property of the boundary modes is a direct consequence of the chiral symmetry.

An alternative $\mathcal{P}\mathcal{T}$-symmetric non-Hermitian SSH chain with $d_x(k) = t_1 + t_2 \, {\rm cos}(k)$, $d_y(k) = t_2 \, {\rm sin}(k)$ and $d_z = i \gamma /2$, i.e., $r_R = r_L = -t_1/t_2$ and, thus, $\ket{\psi_R} = \ket{\psi_L}$, where the non-Hermiticity is included as a staggering potential instead of a hopping term \cite{lieu}, was experimentally realized in Ref. \onlinecite{wiemannkremerplotniklumernoltemakrissegevrechtsmanszameit}. This model is special in the sense that it features gapless regions consistent both with analysis of periodic systems as well as with our prediction for the gap closing at $|r_L^* r_R| = 1$, i.e., $t_1 = \pm t_2$.

{\it Two-dimensional chiral states.---} To further emphasize the generality of our approach, we study the two-dimensional Rice-Mele model \cite{ricemele} with non-Hermitian hopping terms of which a different version is studied in Ref.~\onlinecite{wangzhangsong}. This model exhibits a non-Hermitian analogue of a Chern insulator phase with chiral edge states. The Bloch Hamiltonian is given by Eq.~(\ref{eqgenblochham}) with
\begin{eqnarray}
d_x ({\bf k}) &=& t_+ (k_x) + t_-(k_x) \, {\rm cos}(k_y),\nonumber \\
d_y ({\bf k}) &=& t_-(k_x) \, {\rm sin}(k_y) + i \frac{\gamma}{2}, \,\,\, d_z ({\bf k}) =  t'(k_x),
\end{eqnarray}
where $t_\pm (k_x) = t_1 \pm \delta \, {\rm cos}(k_x)$ and $t'(k_x) = - \Delta \, {\rm sin}(k_x)$. The sites $A$ and $B$ from the previously discussed 1D system now correspond to 1D chains, which together form a 2D lattice. In the periodic bulk EPs occur for ${\rm sin}(k_x) = \pm \sqrt{(\gamma^2/4 - 4t_1^2)/\Delta^2}$ at $k_y = 0$ and ${\rm sin}(k_x) = \pm \sqrt{(\gamma^2/4 - 4\delta^2)/(\Delta^2- 4\delta^2)}$ at $k_y = \pi$. For open boundary conditions in the $y$ direction the exact wave-function solutions are given in Eqs.~(\ref{eqexactwavefctsolright}) and (\ref{eqexactwavefctsolleft}) with $r_R = - [t_+ (k_x)  - \gamma/2]/t_- (k_x)$ and $r_L = - [t_+ (k_x) + \gamma/2]/t_- (k_x)$, such that Eq.~(\ref{eqimportantsolv}) [or, equivalently, Eq.~(\ref{eqimportant})] yields 
\begin{equation}
{\rm cos}(k_x) = \frac{\gamma^2}{16 t_1 \delta}, \, \, \, \pm \sqrt{\frac{\gamma^2/8 - t_1^2}{\delta^2}} \ ,\label{cherncrit}
\end{equation}
where $P$ jumps.

\begin{figure}[t]
  \centering
  {\includegraphics[width=0.99\columnwidth]{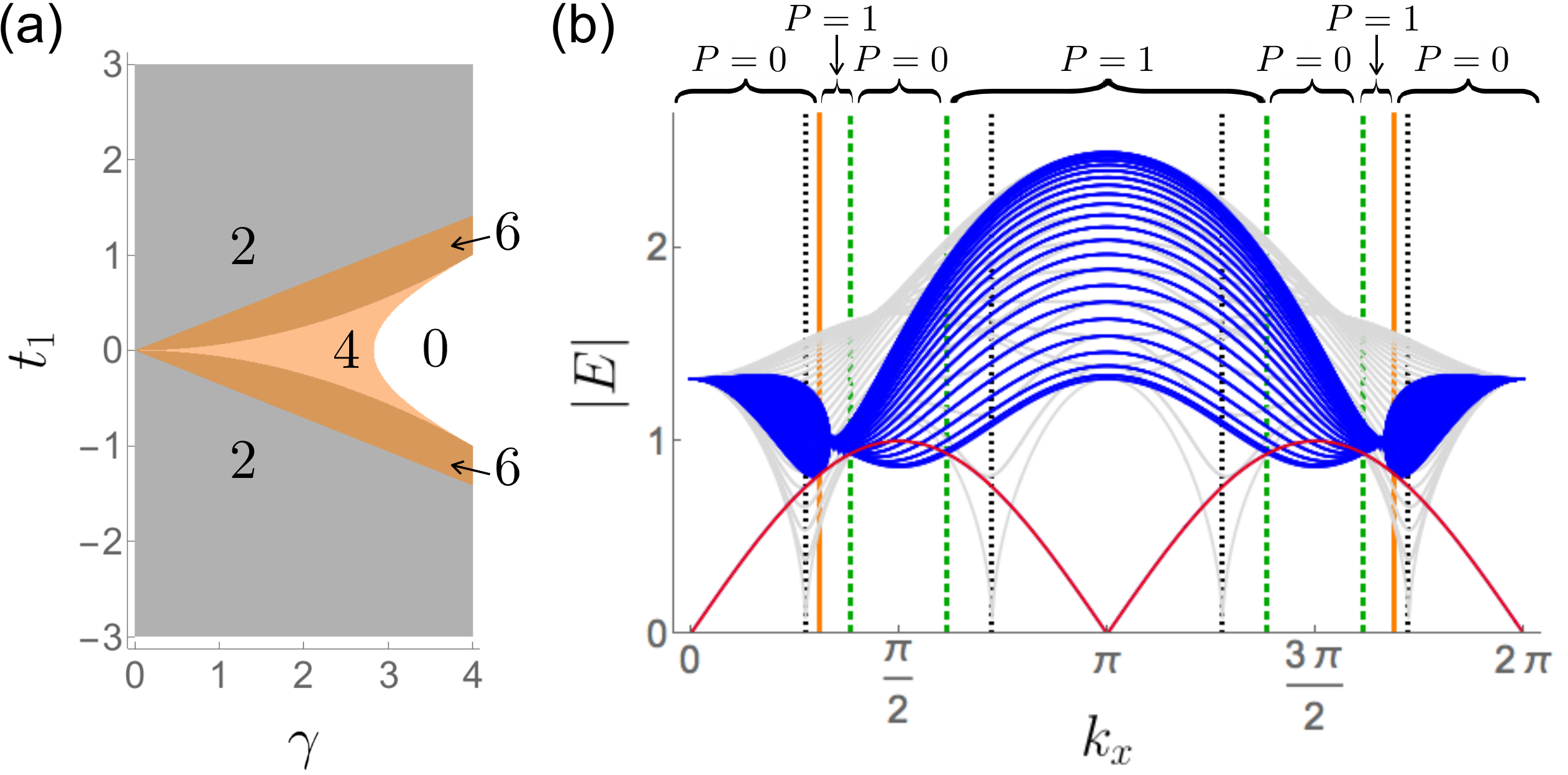}}
  \caption{Non-Hermitian Chern insulators. (a) The bulk phase diagram at $\delta=\Delta=1$ where the different regions are labeled by the number of solutions of Eq.~(\ref{cherncrit}). Chiral edge states occur when this number is nonzero. (b) The absolute value of the energy spectrum---periodic in gray and open in blue and red---of the non-Hermitian Rice-Mele model as a function of $k_x$, with $t_1 = \Delta = \delta = 1$, $\gamma = 3$, and $N=46$. The open system has odd length with a broken unit cell at $n=46$. $P$ jumps precisely when the chiral mode (red) enters the bulk spectrum (blue) but away from the periodic EPs.} 
  \label{fig_energy_semiricemele}
\end{figure}

Saliently, the (non)existence of solutions in Eq.~(\ref{cherncrit}) provide key information about the phase diagram of the open system---only when such solutions exist, there are chiral edge states connecting the valence and conduction bands leading to the phase diagram in Fig.~\ref{fig_energy_semiricemele}(a). It follows that the region without solutions, labeled $``0"$ in the phase diagram, is always separated from the other regions by closing of the bulk energy gap. These simple but striking conclusions are fully supported by our numerical simulations \cite{footnote}. 

The energy spectrum for a system with an odd number of $A$ and $B$ chains and a choice of parameters yielding six solutions to Eq.~(\ref{cherncrit}) is displayed in Fig.~\ref{fig_energy_semiricemele}(b). It is evident that, again, the open and closed systems are qualitatively different: the periodic system is semimetallic with bulk gap closings as a function of $k_x$ (shown in grey) while the open system bulk spectrum (blue) remains gapped and is accompanied by the exactly solvable chiral edge state with energy $E(k_x)=d_z(k_x) = - \Delta \, {\rm sin}(k_x)$ shown in red. The biorthogonal polarization, $P$, reveals that the right (left) mover is localized to the edge $n=1$ ($n=N$) and accurately predicts the six $k_x$ values where the chiral band merges with the bulk via Eq.~(\ref{cherncrit}). If we instead consider a model with an even number of $A$ and $B$ chains, there is both a left and right mover in the region where $P=1$ and no states in the gap when $P=0$ in accordance with our general prediction \cite{footnote}.

{\it Stability and domain wall zero-modes. ---} We now address the stability of the discussed open boundary physics. For concreteness, we focus our discussion on the non-Hermitian SSH model in Eq. (\ref{eqsshham}). 
As mentioned before, the {\emph{bulk}} spectrum behaves discontinuously when going from closed to open boundary conditions in the thermodynamic limit. This can be intuitively understood from the non-Hermitian skin effect, meaning that a macroscopic number of eigenmodes pile up exponentially close to the boundary. 
More formally, by introducing a term $\Gamma$ that rescales the hopping between the ends of the chain, such that the chain is open when $\Gamma = 0$ and periodic when $\Gamma =1$ \cite{xiong}, we find that there is an exceptional point at $\Gamma=0$ in the thermodynamic limit and that in finite systems there is a characteristic scale $\Gamma_c \propto e^{- \alpha N}$, with $\alpha\approx 0.58$ for the parameters used in Fig. \ref{fig_energy_semissh}, which marks the crossover to the periodic boundary physics.
Since $\Gamma_c$ is exponentially small in the system size \cite{footnote}, it is natural to ask how relevant the open boundary physics actually is for realistic setups, and for domain walls between different phases rather than sharp boundaries.
This is a subtle issue since the coupling between the two ends in an experimental setup also is naturally exponentially small in the separation between ends. 
Thus it becomes a question of competing energy scales whether the qualitatively different open or closed scenario, each predicting boundary modes in a quite different parameter regime, describes the system. 
To understand this crossover microscopically, we consider a system in ring geometry with two distinct domains $\alpha$ and $\beta$, with $\alpha$ consisting of $N_{\alpha}$ unit cells representing the non-Hermitian model of interest, and $\beta$ with $N_{\beta}$ unit cells corresponding to a trivial Hermitian chain that is gapped and has no zero modes. 
In this scenario, the end modes of the non-Hermitian chain enter the Hermitian chain with an exponentially decaying tail, whose width is set by the gap in the spectrum of the Hermitian chain, and which defines an effective amplitude $\tau_\beta$ for the domain-wall zero mode of leaking through domain $\beta$.
 If the gap in the spectrum of domain $\beta$ is large enough, $\tau_\beta < \Gamma_c$, and the open system physics prevails. 
 However, if domain $\beta$ is more penetrable such that $\tau_\beta > \Gamma_c$, the zero modes can pass through the Hermitian chain and periodic physics is restored \cite{footnote}. 
 Remarkably, this provides a simple experimental knob to tune the system through the exceptional point at $\Gamma_c$. 

{\it Concluding discussion.---} In this work, we have systematically addressed the issue of qualitative changes to the bulk-boundary correspondence in non-Hermitian models, where closed and open systems can exhibit glaringly different bulk spectra. In particular, we introduced and studied with several concrete examples a new quantity coined the biorthogonal polarization [see Eq.~(\ref{eqbiorthpol})] that predicts band-touching points in systems with {\emph{open}} boundaries, where the number of boundary or domain-wall zero modes changes. To illustrate our general conclusions we have studied lattice models with exactly solvable boundary modes, and explicitly shown that the biorthogonal polarization $P$ exhibits a jump precisely when boundary modes are attached to or detach from the bulk spectrum. These transitions in the open system are found to generally decouple from the individual properties of the left and right eigenstates, as well as of the Bloch bands in the closed periodic systems, leading to strikingly different phase diagrams. Furthermore, effectively interpolating between open and periodic boundary conditions in closed geometries with two inequivalent domains, we have achieved a conclusive understanding of the subtle crossover between open boundary and periodic boundary physics. 

Recently, various bulk topological invariants have been defined for non-Hermitian Bloch Hamiltonians of periodic systems, some also using the framework of biorthogonal quantum mechanics \cite{lee,shenzhenfu, lieu, yinjiangliluchen,wangzhangsong,leykambliokhhuangchongnori,esakisatohasebekohmoto, rudnerlevitov}. However, boundary modes are found in parameter regimes that can drastically differ from the predictions of such bulk topological phase diagrams of periodic systems. Progress towards addressing these issues was very recently reported in Refs. \onlinecite{yaosongwang, yaowang}, which put forward an {\em ad hoc} hybrid construction of quantities obtained by supplementing the Bloch states with information numerically extracted from the open boundary eigenstates. Our present analysis settles this controversial discussion of the bulk-boundary correspondence in non-Hermitian systems by first providing a universal quantity that accurately predicts bulk transitions in systems with open boundaries that are associated with the (dis)appearance of boundary modes. 

\acknowledgments
{\it Acknowledgments.---} We would like to thank E. Ardonne for useful discussions, and G. van Miert and M. Trescher for related collaborations. F.K.K., E.E., and E.J.B. are supported by the Swedish research council (VR) and the Wallenberg Academy Fellows program of the Knut and Alice Wallenberg Foundation. J.C.B. acknowledges financial support from the German Research Foundation (DFG) through the Collaborative Research Centre SFB 1143.


\begin{thebibliography}{10}
\bibitem{otherapproach}
We note that two of the four solutions in Eq.~(\ref{eqresultopenssh}) predicting band attachment in the even chain were found in Ref.~\onlinecite{yaowang} using a more involved approach.
\end{thebibliography}
\end{document}